\documentclass{elsart3p}
\usepackage{amsmath,amssymb,bm,graphicx,epsfig,psfrag,ulem}

\newcommand{\Rey}{\mbox{\rm Re}}            

\begin{document}

\begin{frontmatter}

\title{Stretching in a model of a turbulent flow}

\author[Newcastle]{Andrew~W.~Baggaley\corauthref{cor}},
\ead{a.w.baggaley@ncl.ac.uk}
\author[Newcastle]{Carlo~F.~Barenghi} and
\ead{c.f.barenghi@ncl.ac.uk}
\author[Newcastle]{Anvar~Shukurov}
\ead{anvar.shukurov@ncl.ac.uk}
\address[Newcastle]{School of Mathematics and Statistics, University of
Newcastle, Newcastle upon Tyne, NE1 7RU, UK}

\corauth[cor]{Corresponding author.}
\journal{Physica D.}


\begin{abstract}
Using a multi-scaled, chaotic flow 
known as the KS model of turbulence \cite{Fung:1992}, we investigate the dependence of Lyapunov
exponents on various characteristics of the flow. We show that the KS model
yields a power law relation between the Reynolds number and the maximum
Lyapunov exponent, which is similar to that for a turbulent flow with the same
energy spectrum. Our results show that the Lyapunov exponents are sensitive to
the advection of small eddies by large eddies, which can be explained by considering 
the Lagrangian correlation time of the smallest scales.
We also relate the number of stagnation points within a flow
to the maximum Lyapunov exponent, and suggest a linear dependence between the two characteristics.
\end{abstract}

\begin{keyword}
\PACS{47.~27.~Eq, 47.~52.~+j, 47.~27.~Gs}
\end{keyword}
\end{frontmatter}

\section{Introduction}
Measures of stretching, such as the Lyapunov exponent, are important tools for
understanding the nature of dynamical systems. For example, the maximum
Lyapunov exponent can provide information about the complexity of an attractor
via the Kaplan-Yorke dimension \cite{Froehling:1981}, or the rate of loss of
information in the system \cite{Wales:1991}. The use of Lyapunov exponents in
turbulent flows is far too great to list here; examples range from probing the
onset of turbulence \cite{Brandstater:1987}, to detecting inhomogeneity in
hydromagnetic convection \cite{Kurths:1991}. Another interesting 
application
arises in dynamo theory; it was shown \cite{Vishik:1989} that Lyapunov
exponents provide an upper bound for the growth rate of a fast dynamo, and also a non-trivial combination of Lyapunov exponents gives an exact growth rate for the small scale turbulent dynamo \cite{Chertkov:1999}.

In this work we use a model turbulent flow, known as the Kinematic Simulation
(KS) model, that has been primarily used as a Lagrangian model of turbulence
\cite{Fung:1992,Fung:1998}. An important feature of KS is that it allows full
control of the energy spectrum; moreover, its simple analytic structure means
that numerical differentiation is not required in calculating the Lyapunov
exponents. The KS model has been shown to be in good agreement with results
obtained from direct numerical simulations (DNS) of turbulent flows, particularly with respect to
Lagrangian statistics such as two-particle dispersion
\cite{Fung:1998,Malik:1999,Osborne:2006}. The use of the model is spreading
rapidly to many other areas such as aeroacoustics and biomechanics. This flow
has also been shown to be a hydromagnetic dynamo \cite{Wilkin:2007}. Motivated
by the success of the KS model and its applications in magnetohydrodynamics,
our aim is to check the agreement between the model and turbulent flows
with respect to
Lyapunov exponents.
It has been shown that the KS model exhibits Lagrangian chaos
\cite{Fung:1998,Malik:1999}; we shall quantify this feature using the largest Lyapunov
exponent.

\section{The velocity field}
    The KS model prescribes the flow velocity at a position
$\mathbf{x}$ and time $t$ through the summation of Fourier modes
with randomly chosen parameters. These
modes are mutually independent, therefore the advection of small eddies by
large eddies is not included in the model. More precisely, the velocity field is prescribed to be \cite{Osborne:2006}
\begin{equation}\label{uF}
{\bf u}({\bf x},t)= \sum_{n=1}^{N}\left({\bf A}_n \times {\bf k}_n
\cos\psi_n + {\bf B}_n \times {\bf k}_n \sin\psi_n \right),
\end{equation}
where $\psi_n={\bf k}_n \cdot {\bf{x}} + \omega_n t$ and $N$ is the number of
modes. The unit vectors $\hat{\mathbf{k}}_n$ are
chosen randomly,
and $\mathbf{k}_n=k_n\hat{\mathbf{k}}_n$ where $k_n$ is the wavenumber of the
$n^{\textrm{th}}$ mode. The construction of $\mathbf{A}_n$ and $\mathbf{B}_n$, which are time independent, is explained in the appendix.
Even though the parameters of the flow are chosen randomly, they do not
necessarily change with time, so the flow is not necessarily random.
We adopt a normalised energy spectrum of the KS flow $E(k)$, which is
a modification of the von K\'{a}rm\'{a}n energy spectrum,
\begin{equation}\label{Ek}
        E(k)=k^4(1+k^2)^{-(2+p/2)}e^{-1/2(k/k_{N})^2},
      \end{equation}
      which reduces to $E(k)\propto k^{-p}$ in the inertial
range $1\ll k\ll k_N$, with $k=1$ at the integral scale; $p=5/3$ produces the
Kolmogorov spectrum.
As mentioned previously, a useful feature of the KS model
is the ability to vary the slope $p$ in the inertial range.
The flow is incompressible
and time dependent; the frequency of the $n^\mathrm{th}$ mode, $\omega _n$ is
inversely proportional to its turnover time,
\begin{equation}
\omega_n = \sqrt{k_n^3E(k_n)}.
\end{equation}
      It is convenient to write the unit vector $\hat{\mathbf{k}}_n$ as
         \begin{equation}
           \hat{\mathbf{k}}_n=\left(
           \begin{array}{ c }
             \sqrt{1-\zeta_n^2}\cos \theta_n \\
             \sqrt{1-\zeta_n^2}\sin \theta_n \\
             \zeta_n
           \end{array}
           \right),
         \end{equation}
         where, $\theta_n \in [0, 2\pi)$ and $\zeta_n \in [-1, 1]$,
are uniformly distributed random numbers,  to ensure that $\hat{\mathbf{k}}_n$ are isotropically
distributed. With
         \begin{equation}
           k_n=k_1\left( \dfrac{k_{N}}{k_1}\right)^{(n-1)/(N-1)},
         \end{equation}
     the effective Reynolds number is introduced using the requirement
that the dissipation and eddy turnover times are equal to each other at $k=k_N$,
\begin{equation}
  \Rey=(k_{N}/k_1)^{(p+1)/2}.
\end{equation}
Since the maximum value of $E(k)$ and the integral scale remain unchanged in the
models discussed here, any variation in $\Rey$ can be thought to be caused by 
changes in the fluid viscosity.
Fig.~1 shows the energy spectrum of the KS flow, obtained numerically after
fast Fourier transforming $\mathbf{u}$ calculated from Eq.~(\ref{uF}) on a $128^3$ mesh. We also show a slice, in the $z$ plane, of the corresponding vorticity field, with velocity vectors.

%
\begin{figure*}
   \begin{center}
      \includegraphics[width=0.49\textwidth]{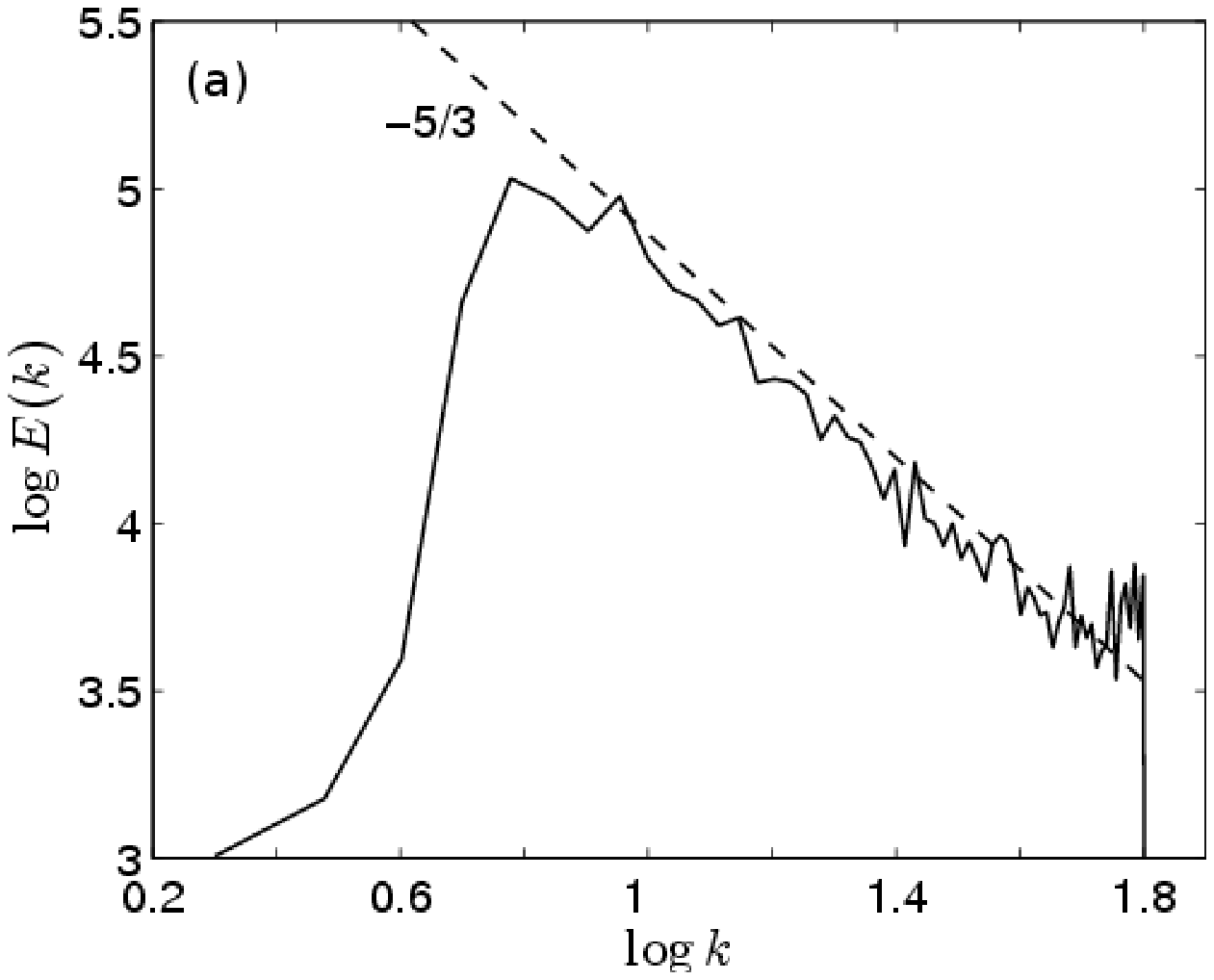}\hfill
      \includegraphics[width=0.38\textwidth]{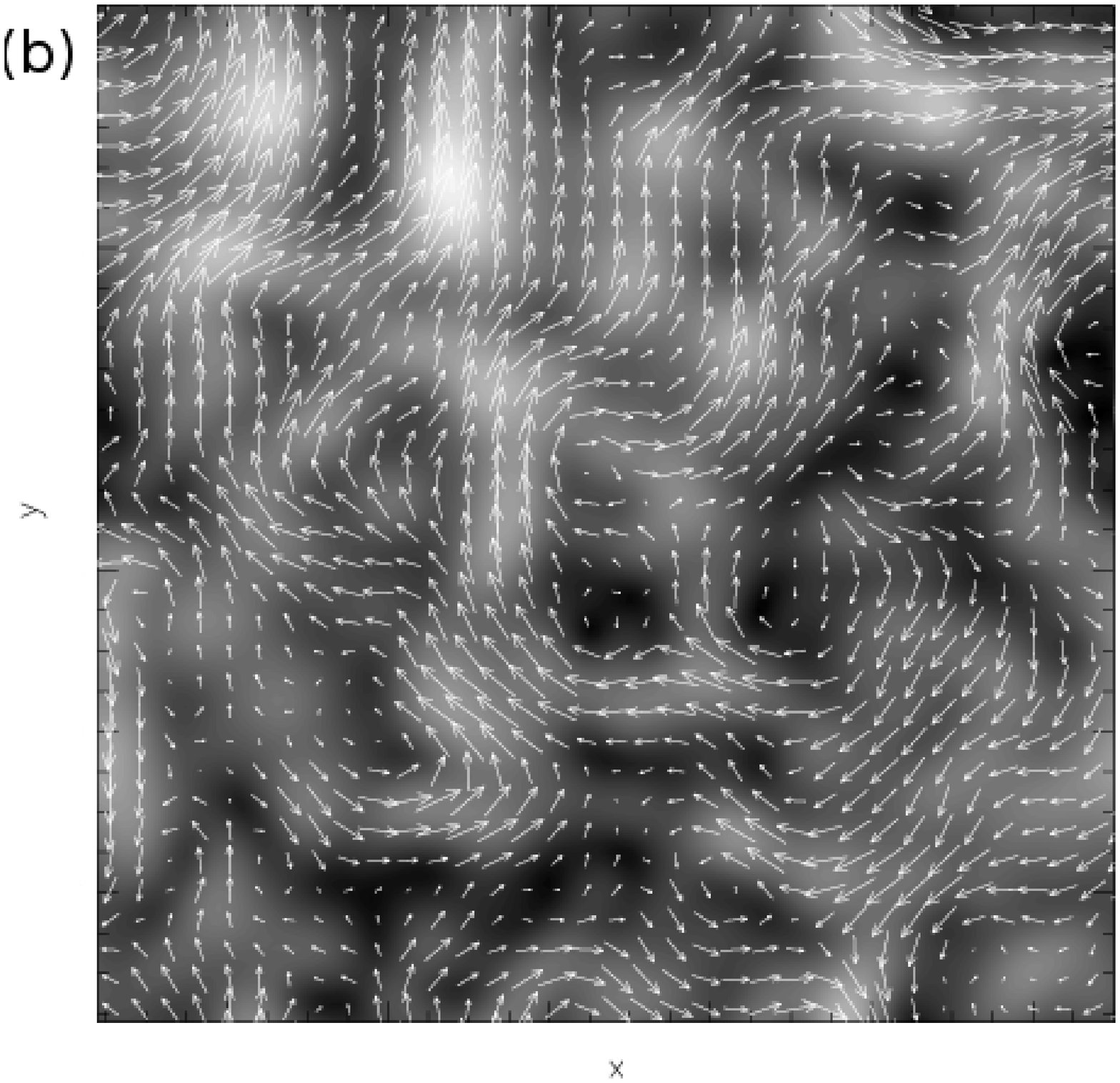}
\end{center}
      \caption{\label{fig1}
     {\bf(a)}  The energy spectrum, $E(k)$, showing the imposed $p=5/3$ slope, 
               as obtained by Fourier transform of Eq.~(\ref{uF}) with $N=20$, 
               $k_1=10$ and $k_N=400$. 
     {\bf(b)}  Slice in the $z$ plane of the vorticity field generated by taking curl of the velocity field from {\bf(a)}, lighter shading indicates higher vorticity. Velocity vectors are shown in white.}
\end{figure*}
\section{The Lyapunov exponents}
\begin{figure}
   \begin{center}
      \includegraphics[width=0.49\textwidth]{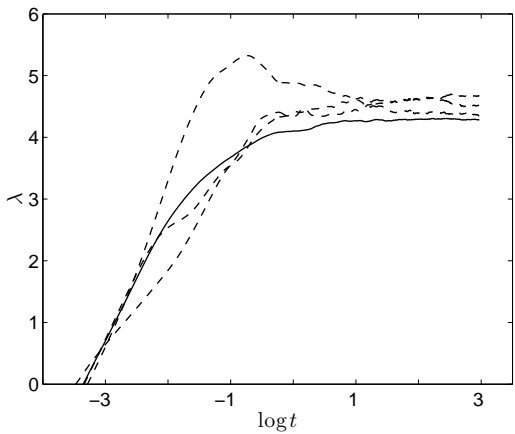}
\end{center}
      \caption{\label{fig2}The average Lyapunov exponent obtained from
    averaging over 500 particles within several realizations of the flow
    (\protect\ref{uF}) is shown with dashed lines. The average over 50
    realisations of the flow is shown with solid curve. The flow has $N=20$,
    $k_1=6$ and $k_N=90$.}
\end{figure}

To obtain the spectrum of Lyapunov exponents, $\lambda_i$, we measure the average
rates of exponential divergence of nearby fluid particle trajectories. If the system is
chaotic, at least one Lyapunov exponent is positive. The procedure to calculate
    the Lyapunov exponents consists of monitoring the evolution of an
    infinitesimal fluid sphere moving with the flow. The sphere,
    deformed by the flow, rapidly becomes an ellipsoid. Then the Lyapunov exponents
    are defined as
\begin{equation}
  \lambda_i=\lim_{t \to +\infty} \frac{1}{t}\log_2 \frac{p_i(t)}{p_i(0)},
\end{equation}
where $p_i(t)$ is the ellipsoid's $i^{\textrm{th}}$ principal axis, and $i=1,2,3$.
    Another feature of the KS flow is that it is time reversible (unlike `real' turbulence), therefore the second Lyapunov exponent vanishes \cite{Benettin:1976,Bec:2006}. 
    We now consider two remaining exponents, which must have opposite signs, $\lambda_1=- \lambda_3$, since the
    flow is incompressible ($\nabla \cdot \mathbf{u}=0 $), the sum of the Lyapunov exponents must be zero. Hence we only need
    to calculate one exponent, $\lambda=\max (\lambda_i)$.
    Following Wolf et al.\ \cite{Wolf:1985}, consider a sphere whose centre, at $\mathbf{x}_0$, moves along
    a trajectory defined by
\begin{equation}\label{dxdt}
  \dfrac{d}{dt}\mathbf{x}_0(t)=\mathbf{u}(\mathbf{x}_0,t),
\end{equation}
with $\mathbf{u}$ obtained from Eq.~(\ref{uF}). As the sphere follows a trajectory in the flow, its shape is deformed to an ellipsoid by stretching and compression. To the linear approximation
    in the sphere radius $\eta$, $\mathbf{x}_0$ remains the centre of the deformed
    ellipsoid.  Positions of the points on the
    surface of the sphere $\boldsymbol{\eta}=\mathbf{x}-\mathbf{x}_0$, where
    $\mathbf{x}_0$ is the position of the centre of the ellipsoid, obey the linearised  equations of motion

\begin{equation}\label{eta}
  \dfrac{d}{dt}\eta_i(t)=D_{ij}\eta_j,
\end{equation}
where $D_{ij}=\partial u_i / \partial x_j$ and the summation convention is assumed.
    We integrate Eq.~(\ref{dxdt}) and (\ref{eta})
    numerically, normalising $\boldsymbol{\eta}$ at regular intervals as to keep
    the linearisation valid. We then
    take the temporal average of the magnitude of $\boldsymbol{\eta}$ to
    recover the maximum Lyapunov exponent. Finally, we average the results over 500
    particles to improve statistics. Since detailed behavior of $\lambda$ can vary
significantly between different realisations of the flow, we further take an
    ensemble average over 50 different realisations of the KS model with the
    same non-random parameters. The results of one such run are shown in Fig.~\ref{fig2}.
    Before beginning the simulations, the code was tested by computing Lyapunov 
    exponents for some well known chaotic flows \cite{Sprott:2003}. 


\section{Scaling of the Lyapunov exponent with the Reynolds number}
We now determine how $\lambda$ scales with the properties of the flow. We
    begin by considering the relationship between the maximum Lyapunov
    exponent, $\lambda$, and the Reynolds number, $\Rey$. Following Ruelle
    \cite{Ruelle:1979}, we increase $\Rey$ by introducing smaller scales (i.e.,
    increasing $k_N$), keeping the same number of modes
    and the same $k_1=1$. In this way we introduce motions with higher velocity shear rate. 
    The maximum Lyapunov exponent is the modulus of the long-term average of the velocity gradient 
    in the Lagrangian frame.
    In fully developed turbulence this is related to the turnover time of the 
smallest eddies, where the turnover time of an eddy of size $\ell$ is
\begin{equation}
  \tau(\ell) \sim \tau_L\left(\dfrac{\ell}{L}\right)^{1-h},
\end{equation}
where $L$ is the integral scale, $U$ the corresponding velocity (with
    $\tau_L=L/U$) and $h$ the H\"{o}lder exponent of the velocity
    field, introduced as
\begin{equation}\label{hoelder}
  u(\ell) \sim U\left(\dfrac{\ell}{L}\right)^h.
\end{equation}
Then $\lambda$ scales with $\Rey$ as
\begin{equation}\label{Realpha}
  \lambda
        \sim \dfrac{1}{\tau_\eta}
        \sim \dfrac{1}{\tau_L}\left( \dfrac{\ell_{\eta}}{L} \right)^{h-1}
        \sim \dfrac{1}{\tau_L}\Rey^\alpha,
\end{equation}
where $\alpha=(1-h)/(1+h)$
    and $\ell_\eta=2\pi/k_N$ is the Kolmogorov length scale.
    For the Kolmogorov spectrum, $p=5/3$, we have $h=1/3$ and $\alpha=1/2$.
\begin{figure*}
   \begin{center}
      \includegraphics[width=0.49\textwidth]{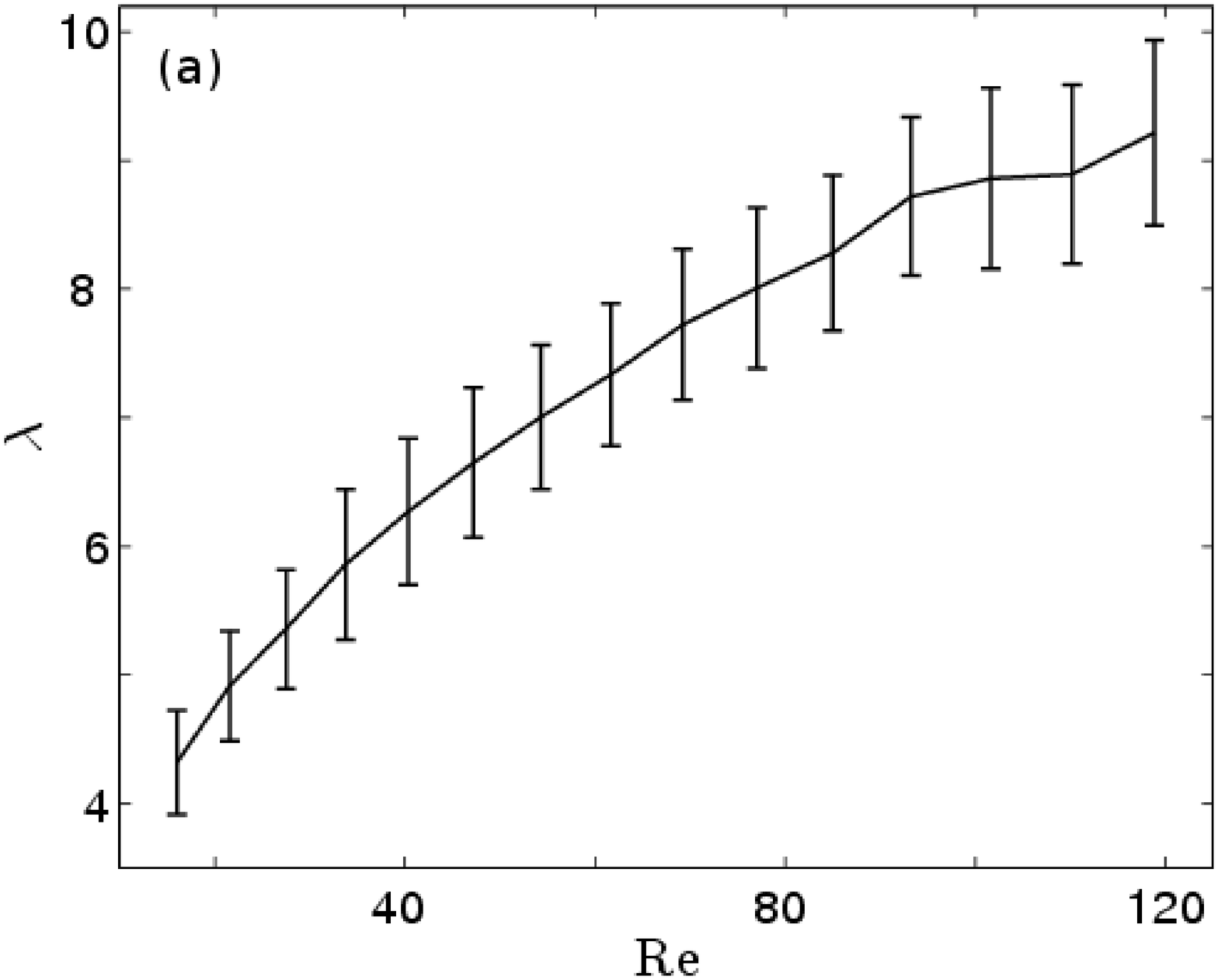}\hfill
      \includegraphics[width=0.49\textwidth]{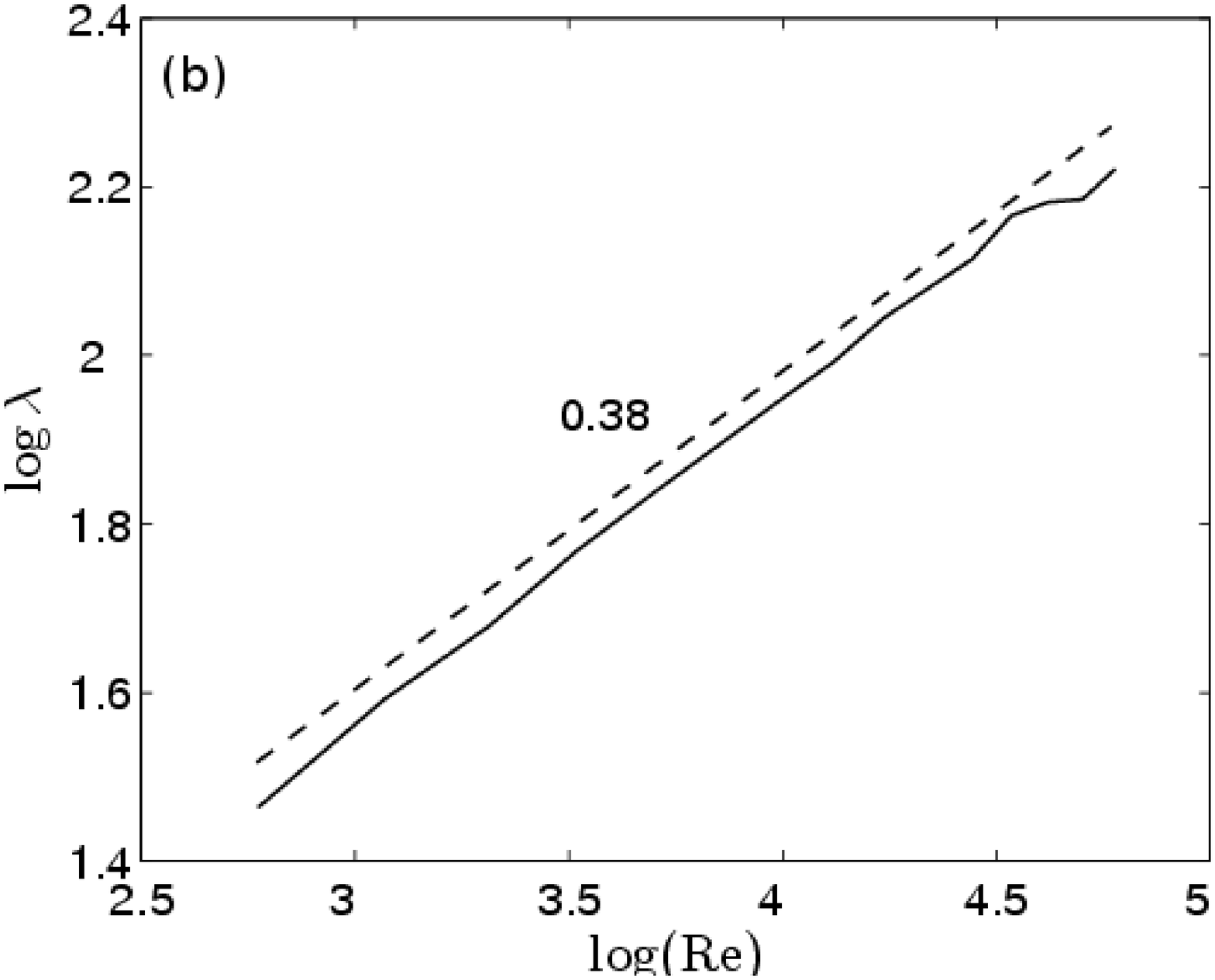}
\end{center}
      \caption{\label{f3}
     {\bf(a)} $\lambda$ against $\Rey$ with error bars, with $N=40$ in
    each realization and $p=5/3$. Errors are calculated from the scatter in
     $\lambda$ between different trajectories in each realisation, and
    between different realisations.
    {\bf(b)}  As in {\bf(a)}, but with the line of best fit shown dashed,
    giving $\alpha \approx 0.38$.}
\end{figure*}

Our numerical simulations, illustrated in Fig.3, show that the  largest Lyapunov exponent of the KS
model scales as $\lambda \propto \Rey^{0.38}$, thus $\alpha$ is smaller than
    Ruelle's prediction. This difference may arise from the lack of
    sweeping of the small eddies by the large eddies in the KS
    model.
    In KS, like in real turbulence, velocity is determined by the large scale `eddies', and velocity gradients are determined by 
    the small scale motions. But, the Lyapunov exponent is related to velocity gradients in the Lagrangian frame, hence the lack of 
    advection of the small eddies is important.
    Because of the lack of advection, the Lagrangian correlation time of eddies of size $1/k_n$ is $1/k_n$, not $1/\omega_n \sim 1/k_n^{3/2}$. The maximum Lyapunov exponent is related to this correlation time, hence there is a reduction in the expected value of $\lambda$, that grows with $\Rey$. 
We are grateful to an anonymous referee who suggested this explanation. 

Hence we must be careful when applying the KS model to area's where stretching is important, especially if we are comparing results with DNS, and scaling with $\Rey$. Dynamo action is one of such areas.

\begin{figure}
   \begin{center}
      \includegraphics[width=0.49\textwidth]{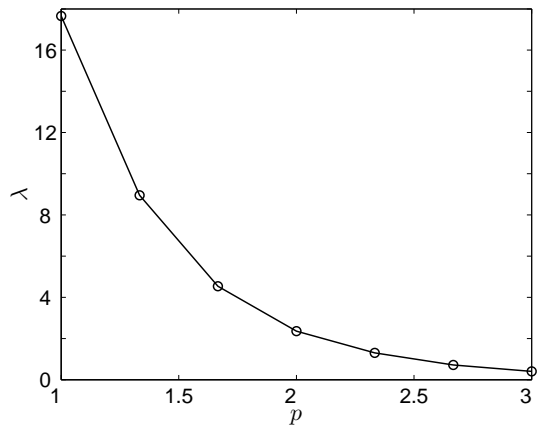}
\end{center}
      \caption{\label{f4} The maximum Lyapunov exponent $\lambda$
    plotted against the spectral slope $p$.}
\end{figure}

\begin{figure}
   \begin{center}
      \includegraphics[width=0.49\textwidth]{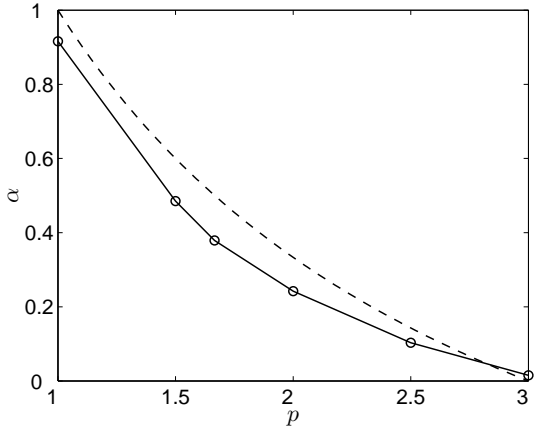}
\end{center}
      \caption{\label{f5} Our estimate of the scaling parameter $\alpha$
    plotted against the spectral slope $p$. The dashed line shows predicted values of $\alpha$ based on Ruelle's work.}
\end{figure}

\section{Lyapunov exponents, spectral slope and stagnation points}
\begin{figure}
   \begin{center}
      \includegraphics[width=0.49\textwidth]{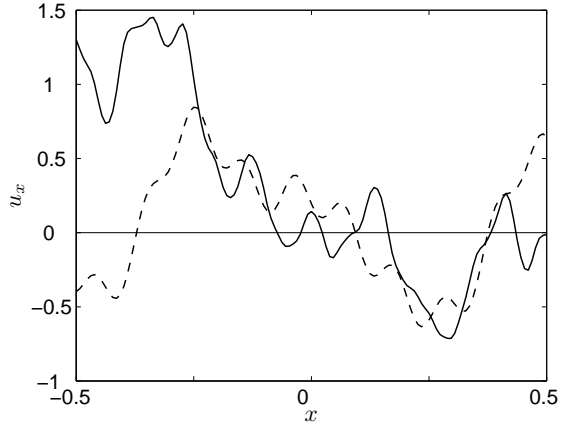}
      \caption{\label{f6} $u_x(x)$ shown for 100 modes in the solid line and and 10 modes in the dashed line. 
      With an increased number of modes the probability of each component being zero at a point increases, hence an
      increase in the number of stagnation points.
A diagnostic of the shear rate, $S=|\frac{\partial v_i}{\partial x_j}\frac{\partial v_i}{\partial x_j}|$ (summation convention assumed) also grows with $N$: $S=11.26 \pm 1.29$ and $17.75 \pm 0.44$, for N=10 and 100, respectively (averaged over 500 realisations).}
\end{center}
\end{figure}

Since the shear rate increases with the wave number for a sufficiently
    steep spectrum, the main cause of stretching are the small scale motions,
    described well by the KS model. Therefore, we should not be too worried
    about the discrepancy described at the end of the previous section. Of our
    primary interest here (motivated by the saturation of the fluctuation dynamo) 
    is the effect of the slope of the energy spectrum plays on stretching.
    Cattaneo et al.\ \cite{Cattaneo:1996} showed, using a simple chaotic velocity field, that
    as a dynamo saturates, Lagrangian chaos in the flow is suppressed. With a complex
    multi-scaled flow such as turbulence this effect would remove energy from small scales first,
    stimulating us to study the effect of steepening the spectrum on $\lambda$.  
    To study how $\lambda$ depends on
    $p$, we fix ${\bf k}_n$ and $k_n/k_1$ and
change the spectral slope $p$ in Eq.~(\ref{Ek}). The resulting values of
$\lambda$, shown in Fig.~\ref{f4}, have been obtained by averaging over $500$
particles in the same flow. These results confirm that the largest Lyapunov
    exponent is controlled mainly by the smaller scales: as less energy is
    given to the small scales, $\lambda$ decreases. The next logical step is
    to investigate the effect of $p$ on the scaling parameter $\alpha$. Indeed,
$h={\textstyle\frac{1}{2}}(p-1)$ in Eq.~(\ref{hoelder}) and then
    \begin{equation}
      \lambda\propto\Rey^{(3-p)/(1+p)},
    \end{equation}
 from Eq.~(\ref{Realpha}). This
dependence is shown dashed in Fig.~\ref{f5} along with the results from our simulations.
As before, our estimates of $\alpha$ are smaller, but we expect this due to the absence of
sweeping. If KS is to be used in areas sensitive to stretching, we suggest the inclusion of 
the advection of small eddies \cite{Fung:1992}. We would then expect the value of $\alpha$ to be
closer to Ruelle's prediction.
It can be expected that the velocity shear rate, and hence the amount of
stretching, is maximum near stagnation points where velocity changes most rapidly in space. 
The number of stagnation points per unit volume (and hence the total magnitude of velocity shear) is sensitive to the number of modes in the KS model. 
The smaller is the value of $k_{n+1}-k_n$ for large $n$, the more numerous will be the stagnation points \cite{Priv:2007}, see Fig.~\ref{f6}. 
Although our velocity field is time-dependent, it has infinite correlation time, so that the density of stagnation points varies little in time.
We computed the maximum Lyapunov exponent for fixed $k_N/k_1$ and $p$, but with
an increasing number of modes, $N$.
%
%
\begin{figure}[h]
   \begin{center}
      \includegraphics[width=0.49\textwidth]{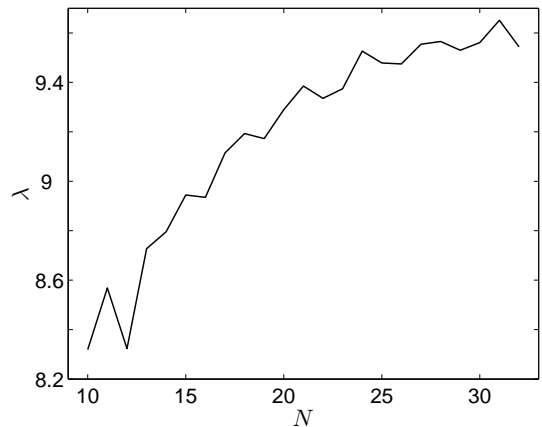}
\end{center}
      \caption{\label{f7}Plot of  $\lambda$ against $N$, showing
saturation of $\lambda$ as $N$ increases}
\end{figure}
\begin{figure}
   \begin{center}
      \includegraphics[width=0.49\textwidth]{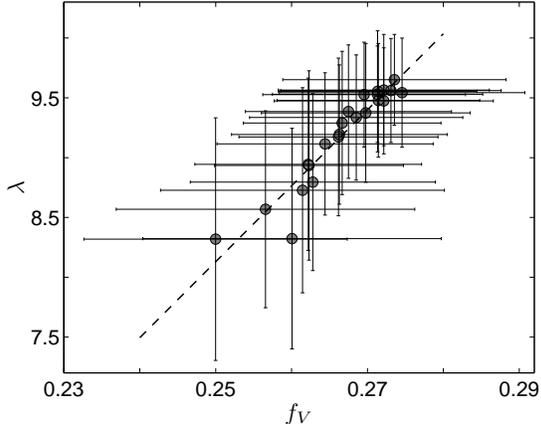}
\end{center}
      \caption{\label{f8} Error bar plot of $f_V(\ell_{\tau})$ against
$\lambda$ with line of best fit, $ \lambda \approx
63.5f_V(\ell_{\tau})-7.7$, suggesting a linear relationship. Calculating the
error in $\lambda$ is explained in Fig. 3, for error in $f_V$ we take error
between realisations.}
\end{figure}
Since the number of stagnation points in the flow increases with $N$,
we expect that $\lambda$ increases too. Fig.~\ref{f7} confirms this and
also shows that $\lambda$ saturates as $N$ grows. The abundance
of stagnation points in the flow can be quantified using what we call the
volume filling factor $f_V(\ell_{\eta})$ calculated as
\begin{equation}
f_V(\ell_{\eta})
=\frac{1}{N_b^3}\sum_{i,j,k=1}^{N_b}
H(0.05 u_\mathrm{rms}-|\mathbf{u}(x_i,y_j,z_k)|),
\end{equation}
where the summation is extended over all the mesh points in the computational
box, $N_b$ is the number of mesh points along each direction, $u_\mathrm{rms}$
is the root-mean-square velocity,  and $H$ is the Heaviside function,
\begin{equation}
  H(x)=\left\{\begin{array}{cc}0,&\mbox{ if }
x< 0,\\1, & \mbox{ if } x\geq 0.\end{array}\right.
\end{equation}
Thus defined, this the fractional volume of the region where $|u|\leq0.05
  u_\mathrm{rms}$. We calculated $f_V(\ell_{\eta})$ in a KS flow with $10 \leq
N \leq 32$ modes with $N_b=128$ in a box whose size is $2\pi/k_1$,
the largest scale in the flow, and we average over 500 realisations of the
flow. Fig.~\ref{f8} shows a scatter plot of $\lambda$ versus $f_V(\ell)$
for an increasing number of modes in the model. Despite the
large error bars, the data suggests a linear relationship. The magnitude of
the errors may be due to the variation between different realisations of the
flow.

\section{Conclusions}
Our simulations confirm that the KS model reproduces
reasonably well the stretching properties of turbulent flows, including the
scaling of the maximum Lyapunov exponent with the Reynolds number despite the
fact that the latter can only be introduced formally in the KS model. The
model values of the Lyapunov exponent are always smaller than theoretical
predictions for turbulent flows; we attribute this to the lack of
advection of the small eddies by the large eddies in the model. Hence we must be careful when applying the KS model to area's where stretching is important. Finally our results suggest a linear relationship between the number of stagnation points in the flow and $\lambda$.


\appendix
\section{Appendix}
We choose $\mathbf{A}_n$, and $\mathbf{B}_n$ randomly, imposing orthogonality with $\hat{\mathbf{k}}_n$, which gives the required spectrum as
 \begin{equation}
        |\mathbf{A}_n \times \hat{\mathbf{k}}_n|=A_n,
      \end{equation}
we proceed in the same fashion for $\mathbf{B}_n$. We then choose
      \begin{equation}
        A_n=B_n=\sqrt{\dfrac{2E(k_n)\Delta k_n}{3}}.
      \end{equation}
This ensures
      \begin{equation}
        \dfrac{1}{V}\int_V\frac{1}{2}|\mathbf{u}|^2 dV=\int_0^{\infty} E(k)dk \sim \sum_{n=1}^{N_k} E(k_n) \Delta k_n,
      \end{equation}
where $\Delta k_n$ is given by
      \begin{equation}
        \Delta k_n=\left\{
        \begin{array}{ c c }
           \dfrac{k_2-k_1}{2}, & n=1, \\
           \dfrac{k_{n+1}-k_{n-1}}{2}, & 1<n<N, \\
           \dfrac{k_{N}-k_{N-1}}{2}, & n=N.
        \end{array}
        \right.
      \end{equation}
\ack{
We are grateful to C. Vassilicos for useful discussions. The helpful comments of anonymous referees are gratefully acknowledged.
}
\bibliographystyle{elsart-num-sort}
\bibliography{my}

\begin{thebibliography}{10}
\expandafter\ifx\csname url\endcsname\relax
  \def\url#1{\texttt{#1}}\fi
\expandafter\ifx\csname urlprefix\endcsname\relax\def\urlprefix{URL }\fi

\bibitem{Bec:2006}
J.~Bec, L.~Biferale, G.~Boffetta, M.~Cencini, S.~Musacchio, F.~Toschi, Lyapunov
  exponents of heavy particles in turbulence, Physics of Fluids 18~(9) (2006)
  091702.

\bibitem{Benettin:1976}
G.~Benettin, L.~Galgani, J.-M. Strelcyn, Kolmogorov entropy and numerical
  experiments, Phys. Rev. A 14~(6) (1976) 2338--2345.

\bibitem{Brandstater:1987}
A.~Brandstater, H.~L. Swinney, Strange attractors in weakly turbulent
  couette-taylor flow, Phys.~Rev.~A 35 (1987) 2207--2220.

\bibitem{Cattaneo:1996}
F.~Cattaneo, D.~W. Hughes, E.~Kim, Suppression of chaos in a simplified
  nonlinear dynamo model, Phys. Rev. Lett. 76 (1996) 2057--2060.

\bibitem{Chertkov:1999}
M.~Chertkov, G.~Falkovich, I.~Kolokolov, M.~Vergassola, Small-scale turbulent
  dynamo, Phys. Rev. Lett. 83~(20) (1999) 4065--4068.

\bibitem{Froehling:1981}
H.~Froehling, J.~P. Crutchfield, D.~Farmer, N.~H. Packard, R.~Shaw, On
  determining the dimension of chaotic flows, Physica D 3 (1981) 605--617.

\bibitem{Fung:1992}
J.~C.~H. Fung, J.~C.~R. Hunt, A.~Malik, R.~J. Perkins, Kinematic simulation of
  homogeneous turbulence by unsteady random fourier modes, J.~Fluid Mech. 236
  (1992) 281--318.

\bibitem{Fung:1998}
J.~C.~H. Fung, J.~C. Vassilicos, Two-particle dispersion in turbulent-like
  flows, Phys.~Rev.~E 57 (1998) 1677--1690.

\bibitem{Kurths:1991}
J.~Kurths, A.~Brandenburg, Lyapunov exponents for hydromagnetic convection,
  Phys.~Rev.~A 44 (1991) R3427--R3429.

\bibitem{Malik:1999}
A.~Malik, J.~Vassilicos, A lagrangian model for turbulent dispersion with
  turbulent-like flow structure: Comparison with direct numerical simulations
  for two particle statistics, Phys.~Fluids 11 (1999) 1572--1580.

\bibitem{Osborne:2006}
D.~Osborne, J.~Vassilicos, K.~Sung, J.~Haigh, Fundamentals of pair diffusion in
  kinematic simulations of turbulence, Phys.~Rev.~E 74 (2006) 036309.

\bibitem{Ruelle:1979}
D.~Ruelle, Microscopic fluctuations and turbulence, Phys.~Lett.~A 72 (1979)
  81--83.

\bibitem{Sprott:2003}
J.~C. Sprott, Chaos and Time-Series Analysis, Oxford University Press, 2003.

\bibitem{Priv:2007}
J.~C. Vassilicos, private communication (2007).

\bibitem{Vishik:1989}
M.~M. Vishik, Magnetic field generation by the motion of a highly conducting
  fluid, Geophys.\ Astrophys.\ Fluid Dyn. 48 (1989) 151--161.

\bibitem{Wales:1991}
D.~J. Wales, Calculating the rate of loss of information from chaotic time
  series by forecasting, Nature 350 (1991) 485--488.

\bibitem{Wilkin:2007}
S.~L. Wilkin, C.~F. Barenghi, A.~Shukurov, Magnetic structures produced by the
  fluctuation dynamo, Phys.~Rev.~Lett. 99 (2007) 134501--134505.

\bibitem{Wolf:1985}
A.~Wolf, J.~B. Swift, H.~L. Swinney, J.~A. Vastano, Determining lyapunov
  exponents from a time series, Physica D 16 (1985) 285--317.

\end{thebibliography}

\end{document}